\documentclass[aps,prd,twocolumn,noshowpacs,nofootinbib,superscriptaddress,groupedaddress, 
floatfix,superscriptaddress, preprintnumbers]{revtex4-2}

\usepackage[utf8]{inputenc}
\usepackage{dcolumn}
\usepackage{amsmath}
\usepackage{amsfonts}
\usepackage{graphicx}
\usepackage{amssymb}%
\usepackage{array,multirow}
\usepackage[breaklinks,colorlinks=true]{hyperref}
\begin{document}

\begin{flushright}
\end{flushright}


\title{Beautiful Exotica}
\author{Micha\l~Prasza\l owicz}
\email{michal.praszalowicz@uj.edu.pl}
\affiliation{Institute of Theoretical Physics, Jagiellonian University, \\ S. \L ojasiewicza 11, 30-348 Krak\'ow, Poland}

\date{\today}

\begin{abstract}

The Chiral Quark--Soliton Model applied to baryons with one heavy quark predicts new exotic states belonging
to three $\overline{\boldsymbol{15}}$ SU(3) multiplets. Here, we extend previous analysis of charm
exotica to the case of beauty. All model parameters are fixed from the charm sector and from the nonexotic
$b$--baryons. We present predictions for masses of beautiful exotica and discuss the decay widths,
which are either very small or relatively large.
  \end{abstract}

\maketitle

\section{Introduction}

Baryons with one heavy quark $Q=b$ or $c$ can be classified according to the SU(3) structure of the light quarks.
Both the Quark Model and the Chiral Quark--Soliton Model ($\chi$QSM)
\cite{Yang:2016qdz,Kim:2017jpx,Kim:2017khv,Polyakov:2022eub}
predict that the light quarks in heavy baryons belong to 
an SU(3) antitriplet of spin 0 or a sextet of spin 1.
Adding a heavy quark results in an antitriplet of spin 1/2 and two hyperfine split sextets of spin 1/2 and 3/2. This structure
is fully confirmed by experiment \cite{PDG}.

In the $\chi$QSM the SU(3) structure follows from the collective quantization of the soliton rotations
\cite{Guadagnini:1983uv,Mazur:1984yf,Jain:1984gp}. While the resulting representations for
the ground states are the same as in the naive Quark Model, the $\chi$QSM predicts exotic
representations of positive parity that cannot be   straightforwardly constructed in the quark picture.
They correspond to pentaquarks (4 quarks and one antiquark).

In the light sector  a prominent example is the elusive
$\Theta^+=uudd{\bar s}$ state, the lightest member of  $\overline{\bf 10}$ SU(3) representation, 
which is predicted to be light and very narrow \cite{Praszalowicz:2003ik,Diakonov:1997mm,Praszalowicz:2024zsy,Praszalowicz:2024mji}.
It was pointed out in Refs.~\cite{Kim:2017jpx,Kim:2017khv,Polyakov:2022eub} that  the lowest lying exotic SU(3) 
representation in the heavy baryon case is $\overline{\boldsymbol{15}}$.

In fact there exist two exotic $\overline{\boldsymbol{15}}$ light SU(3) multiplets~ \cite {Kim:2017jpx}. 
One, carrying angular momentum $J=1$, leads
to two hyperfine split heavy baryon multiplets, and the second one with $J=0$, which is 180~MeV heavier \cite{Praszalowicz:2022hcp},
leads to a spin 1/2 heavy exotica.
In Ref.~\cite{Praszalowicz:2022hcp} we asked a question whether there is space in the experimental charm spectra
for 45 (or perhaps it is better to say: 18 isospin submultiplets) exotic states? 
Here we repeat the same question for $b$ exotica.

Discussing charm sector
in Refs.~\cite{Kim:2017jpx,Kim:2017khv,Polyakov:2022eub,Praszalowicz:2022hcp}
it was argued 
that two out five  $\Omega^{c 0}$ states,  namely $\Omega^{c 0}(3050)$ and $\Omega^{c 0}(3119)$,
discovered in 2017 by the LHCb Collaboration \cite{LHCb:2017uwr,Belle:2017ext,LHCb:2021ptx},
could belong to the exotic  $\overline{\boldsymbol{15}}$,  mainly due to the very small widths and to
the fact that their
hyperfine splittings are equal to the ones of the ground state sextet.\footnote{Note that the ground state
sextet and exotic $\overline{\boldsymbol{15}}$ belong to the same rotational band, and
therefore should have approximately the same value of the hyperfine splitting.}
For different pentaquark interpretations
see Refs.~\cite{An:2017lwg,Yang:2017rpg,Wang:2018alb,Wang:2017smo}.

As far as the remaining charm pentaquarks are concerned, it was shown in Ref.~\cite{Praszalowicz:2022hcp}
hat the members of multiplets based on ($\overline{\boldsymbol{15}}$,
$J=1$) soliton are very narrow (some hint of this behavior was already discussed in Ref.~\cite{Kim:2017khv}),
and -- on the contrary -- states associated with ($\overline{\boldsymbol{15}}, J=0)$ multiplet are  wide.
Both extremes explain why these states are difficult to detect; it is easy to miss a very narrow or
very broad resonant signal.

In the present paper we extend the analysis of Ref.~\cite{Praszalowicz:2022hcp} to exotica with a $b$ quark.
All model parameters are fixed from the spectra of regular baryons both in $c$ and $b$ sector
\cite{Yang:2016qdz,Kim:2017jpx,Kim:2017khv},
and from charm exotica \cite{Praszalowicz:2022hcp}.
Therefore, predictions for beautiful pentaquarks do not require any extra input. For decays, we use parameter
values from Ref.~\cite{Kim:2017khv}. The conclusions are basically the same as in the charm sector:
pentaquarks in $(\overline{\boldsymbol{15}},J=1)$ are very narrow and in  $(\overline{\boldsymbol{15}},J=0)$ 
very wide.

The paper is organized as follows. In Sec.~\ref{sec:chiQSM} we briefly review the main features of the $\chi$QSM.
In Sec.~\ref{sec:HBmasses} we first derive analytical formulae for the baryon masses and then fix splitting
 parameters as functions of the strange moment of inertia $1/I_2$. After constraining $1/I_2$ we compute all pentaquark 
 masses. Next, in Sec.~\ref{sec:decays}, we discuss and compute decays widths, and finally we conclude in Sec.~\ref{sec:conclusions}.

\section{Baryons in the Chiral Quark--Soliton Model}

\label{sec:chiQSM}

It is beyond scope of the present paper to review the $\chi$QSM in detail. We refer the
interested reader to the original paper by Diakonov, Petrov and Pobylitsa
\cite{Diakonov:1987ty} and  the reviews of
Refs.~\cite{Praszalowicz:2024zsy,Praszalowicz:2024mji,Christov:1995vm,Alkofer:1994ph,Petrov:2016vvl} (and references
therein). 

In the Chiral Quark Model light quarks couple in a chiraly invariant way to the
Goldstone boson fields that arise from the spontaneous breaking of chiral symmetry. Goldstone 
bosons are parametrized by an SU(3) matrix $U[\pi,K,\eta]$ with a coupling to quarks being
the constituent quark mass $M\sim 350$~MeV. For the vacuum configuration, where $U=1$, the system reduces
to three free quarks of mass $M$. On the top of that one has to add explicit  mass
terms related to the  Goldstone boson masses ({\em i.e.} to the current quark masses).
In the present approach this explicit symmetry breaking is treated perturbatively.

The soliton arises when the chiral field $U$ is taken in the form of the so called
{\em hedgehog} Ansatz
 \begin{equation}
 U_{0}\, = \,
\left[ 
\begin{array}{ccc}
\multicolumn{2}{c}{\multirow{2}{*}{$u_0$}}    & 0 \\
\multicolumn{2}{c}{}  &0   \\
 0   & 0    & 1
\end{array} 
\right] .
\label{eq:Usu3}
\end{equation}
where
\begin{equation}
u_0=\exp \left( i\,\boldsymbol{n}\cdot \boldsymbol{\tau}%
\,P(r)\right) \, .
\label{eq:hedg}
\end{equation}
Here $\boldsymbol{n}=\boldsymbol{r}/r$ and the profile function $P(r)$ has to vanish at
$r \rightarrow \infty$.

If $P\equiv 0$, the level structure of the corresponding Dirac equation shows a gap
for $-M<E<M$.
For non zero $P(r)$ the eigenvalue structure changes. The lowest positive energy level
(the valence level) sinks into the gap reducing  energy, while the sea levels are distorted
in a way that  energy increases. Such
distortion interacts with  valence quarks changing their wave functions,
which in turn modifiy the sea until a stable configuration is reached \cite{Diakonov:1988mg}. 
This
configuration is called a \emph{chiral soliton}. Note that this procedure is carried
out in a large $N_c$ limit where we have $N_c$ identical copies of the Dirac structure
described above.

In practical terms, a soliton corresponds to a configuration, which minimizes the aggregate energy
\begin{equation}
M_{\rm sol}=N_c\Big[ E_{\rm val}+\sum_{E_n<0} (E_n-E_n^{(0)})\Big]\, ,
\label{eq:Msol1}
\end{equation}
where the sea energy is computed with respect to the vacuum, and the whole sum is
suitably regularized. The minimization is performed with respect to the profile function $P$ \cite{Diakonov:1988mg}.

What happens when one of the $N_c$ light quarks is replaced by a heavy one? The answer
can be inferred from Ref.~\cite{Goeke:2005fs} where the authors
studied soliton behavior in the limit where the current quark masses $m \rightarrow \infty$. 
It turns out that for sufficiently large $m$ the soliton ceases to exist, and the correct heavy quark limit is achieved.

In the present context, Eq.~(\ref{eq:Msol1}) takes the following form:
\begin{align}
M_{\rm sol}&=(N_c-1)\Big[ E_{\rm val}+\sum_{E_n<0} \left(E_n-E_n^{(0)}\right)\Big]  \label{eq:Msol2} \\
                  &+\Big[ E_{\rm val}(m_Q)+\sum_{E_n<0} \left(E_n(m_Q)-E_n^{(0)}(m_Q)\right)\Big].\notag
\end{align}
As  was argued in Ref.~\cite{Goeke:2005fs}, for large $m_Q$,  the sum
over the sea quarks in the second
line of Eq.~(\ref{eq:Msol2}) vanishes, and $E_{\rm val}(m_Q)\approx m_Q$. One copy of the
soliton ceases to exist; however, the remaining $N_c-1$ quarks still form a stable soliton.
The total energy reads therefore
\begin{equation}
M_{\rm sol}=M'_{\rm sol}+ m_Q \, ,
\end{equation}
where the  prime refers to the soliton constructed from $N_c-1$ quarks. In the following we will skip
the prime.

The soliton constructed in a way described above does not carry
any quantum numbers except for the baryon number resulting from the
valence quarks. It is therefore often referred to as a {\em classical baryon}.
Spin and  SU(3) structures appear when the soliton rotations in space and flavor are quantized~\cite{Adkins:1983ya}.
This procedure results in a {\em collective} Hamiltonian analogous to the one
of a quantum mechanical symmetric top, however, due to the Wess-Zumino-Witten
term \cite{WittenCA,Wess:1971yu} and the form of the {\em hedgehog} Ansatz (\ref{eq:Usu3}),
the allowed Hilbert space is truncated to the representations that contain states of
hypercharge $Y'=N_{\rm val}/3$. For light baryons $N_{\rm val}=3$, these are octet and decuplet of ground state
baryons~\cite{Guadagnini:1983uv,Mazur:1984yf,Jain:1984gp}. 

\begin{figure}[h]
\centering
\includegraphics[width=9.cm]{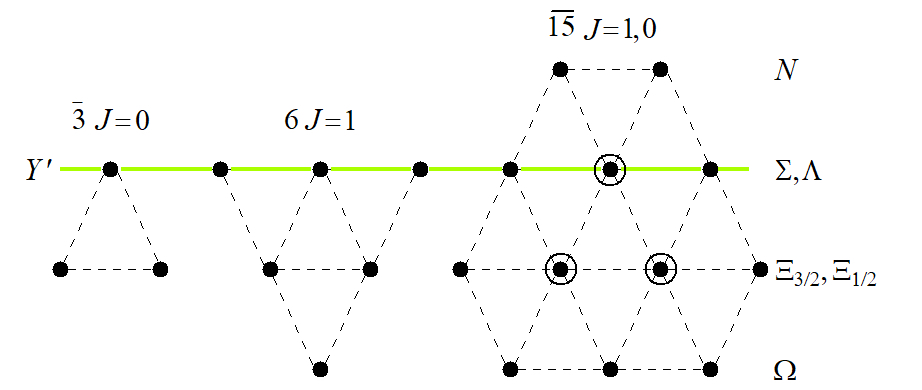} \vspace{-0.2cm}%
\caption{Rotational band of a soliton with one valence quark stripped off.
Soliton spin corresponds to the isospin $T^{\prime}$ of states on the
quantization line $Y^{\prime}=2/3$ (green thick line). We show three lowest
allowed representations: antitriplet of spin 0, sextet of spin 1, and the
lowest exotic representation $\overline{\mathbf{{15}}}$ of spin 1 or 0. 
On the right-hand side we display particle names used
in the present paper.}%
\label{fig:reps}%
\end{figure}

In the case of heavy baryons $N_{\rm val}=2$
(or rather $N_c-1$) and the allowed SU(3) representations  have to contain states of $Y'=2/3$.
These are $\overline{\boldsymbol{3}}$, $\boldsymbol{6}$, and exotic
$\overline{\boldsymbol{15}}$ shown in Fig.~\ref{fig:reps}. They correspond to the
rotational excitations of the meson mean field, which is essentially the same
as for light baryons.

The soliton mass with one heavy quark including rotational energy takes the following form~\cite{Diakonov:2013qta} 
in the chiral limit:
\begin{align}
\mathcal{M}^Q_{\mathcal{R},T'}=& M_{\text{sol}}+m_Q \notag \\
+& \frac{1}{2I_{2}}\left[
C_{2}(\mathcal{R})-T^{\prime}(T^{\prime}+1)-\frac{3}{4}Y^{\prime2}\right] \notag \\
+& \frac{1}{2I_{1}}T^{\prime}(T^{\prime}+1) \, . 
\label{rotmass}%
\end{align}
Here $\mathcal{R}$ refers to the allowed SU(3) representation of the $N_c-1$ light quark sector,
$C_{2}(\mathcal{R})$ stands for the SU(3) Casimir operator. Isospin $T'$ corresponds to the states
with  hypercharge equal to $Y'$ and is equal to the soliton angular momentum~\cite{Guadagnini:1983uv,Mazur:1984yf,Jain:1984gp}, 
which in the following
will be denoted by $J$.
$M_{\text{sol}}\sim N_{c}-1$ denotes the classical soliton mass; $I_{1,2}\sim N_{c}-1$
are moments of inertia. We see that the two lowest representations correspond to the naive quark model.
However, the $\chi$QSM predicts two exotic $\overline{\mathbf{{15}}}$ representations of spin 1 and 0.

Next, we have to include the explicit symmetry breaking Hamiltonian, which  takes the following form~\cite{Blotz:1992pw}:
\begin{equation}
\hat{H}_{\mathrm{{sb}}}=\alpha\,D_{88}^{(8)}+\beta\,\hat{Y}+\frac{\gamma}{\sqrt{3}%
}\sum_{i=1}^{3}D_{8i}^{(8)}\,\hat{J}_{i}, \label{eq:Hsb}%
\end{equation}
where $\alpha$, $\beta$, and $\gamma$ are proportional to the current strange
quark mass (assuming $m_u=m_d=0$) and are
given in terms of the moments of
inertia and the pion-nucleon sigma term.  We shall treat them as free parameters. It is however worth
mentioning that $\alpha$ and $\beta$ are negative by construction, while
$\gamma$ being phenomenologically negative is in fact given as a difference of
two terms of the same order -- see Eq.~(4) in Ref~\cite{Yang:2016qdz}. Furthermore,
$\alpha$  scales as $N_c$, and $\beta$ and $\gamma$
scale as $N_c^0$, however, all of them are numerically of the same order \cite{Praszalowicz:2022hcp}.

Since the collective wave functions of the rotating soliton are known \cite{Diakonov:1997mm},
one can easily compute matrix elements of $H_{\mathrm{{sb}}}$ in the first order of the perturbation
theory \cite{Praszalowicz:2022hcp}:
\begin{align}
\left\langle \hat{H}_{\text{sb}}\right\rangle _{\overline{\boldsymbol{3}},J=0}  &
=\left(\frac{3}{8}{\alpha}+\beta \right) Y_B \equiv \delta_{\overline{\mathbf{{3}}}} Y_B\, , \notag \\
\left\langle \hat{H}_{\text{sb}}\right\rangle _{{\boldsymbol{6}},J=1}  &
=\left( \frac{3}{20}{\alpha}+\beta-\frac{3}{10}\gamma\right) Y_B \equiv \delta_{\mathbf{{6}}} Y_B\, , \notag \\
\left\langle \hat{H}_{\text{sb}}\right\rangle _{\overline{\boldsymbol{15}},J=1}  &
=\left(  \beta+\frac{17}{144}(\alpha-2\gamma)\right)  Y_{B} - (\alpha-2\gamma) \notag \\
&\times \left(  \frac{2}{27}-\frac{1}{24}\left(  T_{B}(T_{B}+1)-\frac{1}{4}Y_{B}^{2}\right)
\right)  \,,\nonumber\\
\left\langle \hat{H}_{\text{sb}}\right\rangle _{\overline{\boldsymbol{15}},J=0}  &
=\left(  \beta+\frac{1}{48}\alpha\right)  Y_{B} \notag \\
&+\alpha \left(  \frac{2}{9}-\frac
{1}{8}\left(  T_{B}(T_{B}+1)-\frac{1}{4}Y_{B}^{2}\right)  \right)  \,
\label{eq:Delta15bar}%
\end{align}
where $Y_{B}$ and $T_B$  denote the hypercharge and the isospin of a given baryon,
respectively. 

There is one more ingredient that we are missing, namely the spin-spin interaction of the heavy quark
with the soliton, which is relevant for sextet and  (${\overline{\boldsymbol{15}},J=1}$). To this end
we add a phenomenological term  corresponding to the chromomagnetic interaction~\cite{Yang:2016qdz}
expressed as
\begin{align}
\hat{H}_{SQ} = \frac{2}{3}\frac{\varkappa}{m_{Q}} \hat{J} \cdot
\hat{S}_{Q} \, ,
\label{eq:ssinter}%
\end{align}
where $\varkappa$ denotes the anomalous chromomagnetic moment that is
flavor independent. The operators ${\hat{J}}$ and ${\hat{S}}_{Q}$
represent  spin operators for the soliton and the heavy quark,
respectively.
This interaction is relevant only for
sextet and  (${\overline{\boldsymbol{15}},J=1}$).
Therefore we arrive at the following
 equations for heavy
 baryon masses
\begin{align}
M_{\mathcal{R}_J,B,s}^{Q}&=
\mathcal{M}_{\mathcal{R},J}^{Q}+\left\langle \hat{H}_{\text{sb}}\right\rangle _{\mathcal{R}, J} \notag \\
&+\delta_{J,1} 
\frac{\varkappa}{m_{Q}}\left\{
\begin{array}
[c]{ccc}%
-2/3 & \text{for} & s=1/2\\
\, & \, & \,\\
+1/3 & \text{for} & s=3/2
\end{array}
\right. \, ,
\label{eq:M3barM6mass}%
\end{align}
where $s$ denotes  spin of a given baryon $B$,  $J$ is the soliton spin,
$\mathcal{M}_{\mathcal{R},J}^{Q}$ is given by (\ref{rotmass}),
$\left\langle \hat{H}_{\text{sb}}\right\rangle _{\mathcal{R}, J}$ by (\ref{eq:Delta15bar})
and $\delta_{J,1} $ is the Kronecker delta.

\section{Heavy baryon masses}
\label{sec:HBmasses}

Before we estimate masses of states in $\overline{\mathbf{{15}}}$ let us
rewrite equation (\ref{rotmass}) for average multiplet masses%
\begin{eqnarray}
\mathcal{M}^Q_{\overline{\mathbf{{15}}},J=0}&=&M_{\mathrm{{{sol}}}}+m_Q+\frac{5}%
{2}\frac{1}{I_{2}}\, , \nonumber \\
\mathcal{M}^Q_{\overline{\mathbf{{15}}},J=1}%
&=&M_{\mathrm{{{sol}}}}+m_Q+\frac{3}{2}\frac{1}{I_{2}}+\frac{1}{I_{1}}.
\label{eq:M15bar}%
\end{eqnarray}
The mass difference
\begin{eqnarray}
\Delta_{\overline{\mathbf{{15}}}}&=&\mathcal{M}^Q_{\overline{\mathbf{{15}}}%
,J=0}-\mathcal{M}^Q_{\overline{\mathbf{{15}}},J=1}=\frac{1}{I_{2}}-\frac
{1}{I_{1}}%
\label{eq:hfsplit}
\end{eqnarray}
is positive, since -- from the estimates of the light sector
\cite{Kim:2017jpx,Diakonov:1997mm} -- $I_{1}\sim(2.5\div3)\times I_{2}$, 
which means that spin 1 soliton is
lighter than the one of spin 0. This was confirmed in the previous study of the
charm sector \cite{Praszalowicz:2022hcp}.

In principle all model parameters entering Eqs.~(\ref{eq:M3barM6mass})
can be calculated in a specific model. However, we follow here the so-called
{\em model-independent} approach introduced in the context of the Skyrme
model in Ref.~\cite{Adkins:1984cf}, where all parameters are extracted from
 experimental data. Below we recapitulate the results of Ref.~\cite{Praszalowicz:2022hcp}.

The average multiplet masses can be conveniently written the following form:%
\begin{align}
{\cal M}_{\overline{\mathbf{{3}}},J=0}^{Q}  &  =m_{Q}+M_{\mathrm{{sol}}}+\frac{1}{2I_{2}%
},\nonumber\\
{\cal M}_{\mathbf{{6}},J=1}^{Q}  &  ={\cal M}_{\overline{\mathbf{{3}}}}^{Q}+\frac{1}{I_{1}%
},\nonumber\\
{\cal M}_{\overline{\boldsymbol{15}},J=1}^{Q}  &  ={\cal M}_{\mathbf{{6}}}^{Q}+\frac
{1}{I_{2}},\nonumber\\
{\cal M}_{\overline{\boldsymbol{15}},J=0}^{Q}  &  ={\cal M}_{\mathbf{{6}}}^{Q}+\frac
{2}{I_{2}}-\frac{1}{I_{1}}={\cal M}_{\overline{\boldsymbol{15}},J=1}^{Q}%
+\Delta_{\overline{\mathbf{{15}}}}. \label{eq:15barav}%
\end{align}
Parameters ${\cal M}_{\overline{\mathbf{{3}}}, J=0}^{Q}$ and $I_{1}$ can be extracted from
the  nonexotic baryons alone~\cite{Yang:2016qdz}. Indeed
\begin{align}
{\cal M}_{\overline{\mathbf{{3}}},J=0}^{Q}=m_{Q}+M_{\mathrm{{sol}}}+\frac{1}{2I_{2}}  &
=\left.  2408.2\right\vert _{c}~,~\left.  5736.2\right\vert _{b}\,,\nonumber\\
{\cal M}_{\mathbf{{6}},J=1}^{Q}={\cal M}_{\overline{\mathbf{{3}}},J=0}^{Q}+\frac{1}{I_{1}}  &
=\left.  2579.7\right\vert _{c},~~\left.  5906.5\right\vert _{b}  \, ,
\label{eq:mqrep}%
\end{align}
where the experimental values in MeV from Ref.~\cite{Yang:2016qdz} have been updated~\cite{Praszalowicz:2022sqx}. We can
compute $I_{1}$ from the mass difference of these two multiplets (in MeV):
\begin{equation}
\frac{1}{I_{1}}={\cal M}_{\mathbf{6},J=1}^{Q}-{\cal M}_{\mathbf{\overline{3}},J=0}^{Q}=\left.
171.5\right\vert _{c}=\left.  170.4\right\vert _{b} \, ,
\label{eq:I1Q}%
\end{equation}
which, as expected, does not depend on the heavy quark mass.
In the following we assume $1/I_1=171$~MeV.

In order to have a handle on $I_{2}$,
and therefore on $\Delta_{\overline{\mathbf{{15}}}}$,
we need to include flavor symmetry breaking (\ref{eq:Hsb}) and a new input from exotica. 
In Ref.~\cite{Yang:2016qdz} the splitting parameters for $\overline{\boldsymbol{3}}$ 
and $\boldsymbol{6}$ were extracted from experiment and read
\begin{align}
\delta_{\overline{\mathbf{{3}}}}=\frac{3}{8}{\alpha}+\beta &
=-180~\mathrm{{MeV}\,,}\nonumber\\
\delta_{\mathbf{{6}}}=\frac{3}{20}{\alpha}+\beta-\frac{3}{10}\gamma &
=-121~\mathrm{{MeV}\,. } \label{eq:d3bard6}%
\end{align}

\begin{figure*}[t]
\centering
\includegraphics[height=8cm]{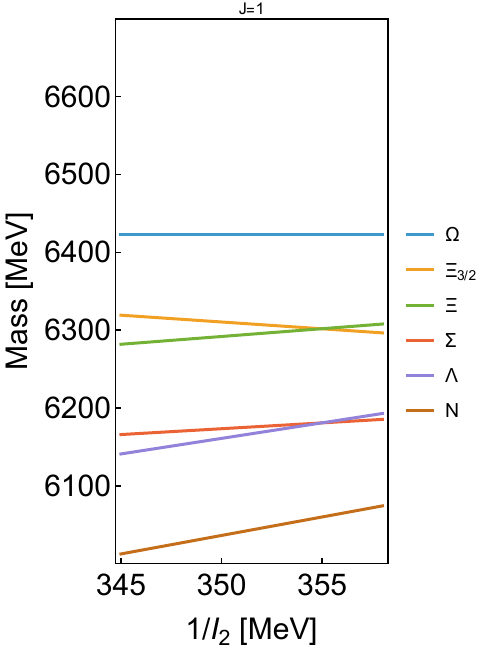} \includegraphics[height=8cm]{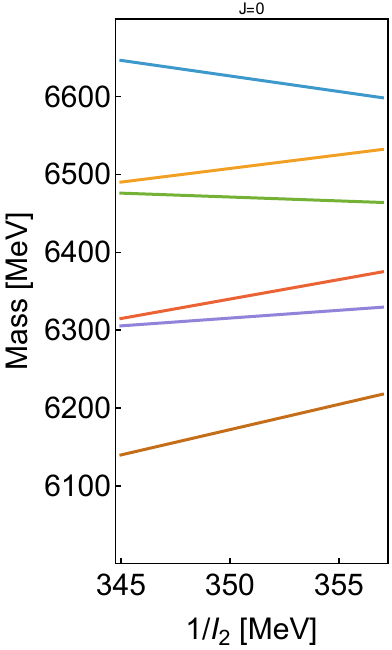} 
\caption{Spectra of exotic beauty multiplets without spin interaction:
${\overline{\boldsymbol{15}},J=1}$ (left) 
and $J=0$
(right) 
in terms of the inverse moment of inertia
$1/I_{2}$.}%
\label{fig:spectra}%
\end{figure*}%

Numerical entries in Eq.~(\ref{eq:d3bard6}) are taken as the average values from Eqs.~(13) and (14) in Ref.~\cite{Yang:2016qdz}.
Note that for the mass splittings in exotica (\ref{eq:Delta15bar}) we need a different combination of
$\alpha$, $\beta$ and $\gamma$, so we cannot predict mass splittings in $\overline{\bf 15}$ from
nonexotic baryons alone. However, if we follow Ref.~\cite{Kim:2017jpx}, where two out of five excited $\Omega^{c}$ hyperons reported by the
LHCb Collaboration in 2017 \cite{LHCb:2017uwr} were interpreted as exotic states
belonging to $(\overline{\mathbf{{15}}},J=1)$, we obtain an additional formula for the spin averaged 
exotic  $\Omega^{c}$ states 
\begin{equation}
\overline{M}_{\Omega,(\overline{\mathbf{{15}}},J=1)}^{c}=3096\;\text{MeV.}
\label{eq:MavOmega}%
\end{equation}
From Eqs.\thinspace(\ref{eq:M15bar}), (\ref{eq:15barav}) and (\ref{eq:Delta15bar}) 
we obtain that%
\begin{align}
\overline{M}_{\Omega,(\overline{\mathbf{{15}}},J=1)}^{c}  &  
={\cal M}_{\mathbf{{6}},J=1}^{c}+\frac{1}{I_{2}}-\frac{1}{6}\left(  \alpha
+8\beta-2\gamma\right)  \, .
\label{eq:MavOmegath}%
\end{align}
Equating
(\ref{eq:MavOmega}) with (\ref{eq:MavOmegath}) together with
Eqs.~(\ref{eq:d3bard6}) gives three independent equations for four parameters
$\alpha$, $\beta$, $\gamma$ and $1/I_{2}$. We solve them in function of
$1/I_{2}$ (in MeV)
\begin{eqnarray}
\alpha &=&  4102.27 - 12 \frac{1}{I_2}\, , \nonumber \\
\beta &=&  -1718.35 + \frac{9}{2} \frac{1}{I_2}\, , \nonumber \\
\gamma &=& -3273.37 + 9\frac{1}{I_2}\, . 
\label{eq:albega}
\end{eqnarray}
Furthermore, we constrain parameter $1/I_{2}$ to the region where  $\alpha$,
$\beta$ and $\gamma$ are negative and of the same order~\footnote{See discussion below Eq.~(\ref{eq:Hsb})}: $1/I_2 = 351 \pm 5$~MeV,
which we use as an estimate of model uncertainty.
This fit and the relation
of the present model parameters to other calculations is in detail discussed
in Ref.~\cite{Praszalowicz:2022hcp}.

Finally we have to include hyperfine interaction (\ref{eq:ssinter}). In Ref.~\cite{Praszalowicz:2022hcp} we estimated
from non-exotic baryons
\begin{align}
\frac{\varkappa}{m_b}
&= \left. 19.4 \right|_{\Sigma_b}
 = \left. 18.8\right|_{\Xi_b}
\label{eq:spintest}
\end{align}
(in MeV). In what follows we assume 19~MeV.

Now, when all parameters are fixed, we can easily predict masses of all exotic $b$--baryons. The results are shown in Fig.~\ref{fig:spectra}
and in Tab.~\ref{tab:mpred}.

\renewcommand{\arraystretch}{1.7}
\begin{table}[h!]
\centering
\begin{tabular}{|c|c|c|c|}
\hline
\multirow{2}{*}{$\overline{\boldsymbol{15}}$}& \multicolumn{2}{c|}{$J=1$} &$ J=0$\\
\cline{2-4}
&$ s=1/2$ &$ s=3/2$ & $s=1/2$\\
\hline
$N^{b} $& 6005--6052 & 6024--6071 & 6146--6211\\
$\Lambda^{b}$ & 6132--6172 & 6151--6191 & 6307--6328\\
$\Sigma^{b}$ & 6155--6170 & 6174--6189 &6320--6370\\
$\Xi_{1/2}^{b} $& 6271--6291 & 6290--6310 & 6475--6465\\
$\Xi_{3/2}^{b}$ & 6305--6287 & 6324--6306 & 6494--6529\\
$\Omega^{b}$ & 6410 & 6429 & 6643--6603 \\
\hline
\end{tabular}
\caption{Mass predictions in MeV for exotic $\overline{\boldsymbol{15}}$ $b-$pentaquarks.
Since two $\Omega^{c}$ states are taken as input, also two $\Omega^{b}$
states have fixed masses that do not depend on $1/I_2$.
\label{tab:mpred}}
\end{table}
\renewcommand{\arraystretch}{1}

Let us observe that assuming $1/I_{2}=351$~MeV and taking $1/I_{1}$
from Eq.~(\ref{eq:I1Q}), we obtain that $(\overline{\mathbf{{15}}},J=0)$
multiplet is heavier from $(\overline{\mathbf{{15}}},J=1)$ multiplet on
average by approximately $180$~MeV. Indeed%
\begin{align}
{\cal M}_{\overline{\boldsymbol{15}},J=1}^{b}  &  \simeq6257.5\;\text{MeV},\nonumber\\
{\cal M}_{\overline{\boldsymbol{15}},J=0}^{b}  &  \simeq6437.5\;\text{MeV.}%
\end{align}
We can also estimate the average mass of the next exotic
representation $(\overline{\boldsymbol{15}}^{\prime}=(p=0,q=4),J=1)$ to be
approximately 6960 MeV, which is indeed substantially heavier than
$\overline{\boldsymbol{15}}$.

\section{Decays}
\label{sec:decays}

\linespread{1.28}
\begin{table*}[t]
\centering
\begin{tabular}{|c|c|c|c|c|}
\hline
\multicolumn{2}{|c|}{decay}&\multicolumn{3}{c|}{$\Gamma$~[MeV]}\\
\hline
$B_1$  & $B_2+\varphi$&  $s_2=\frac{1}{2}$ & $s_2=\frac{3}{2}$ & $\Sigma_{s_2}$\\ 
\hline
\multirow{2}{*}{$\Omega^b({\boldsymbol{\overline{15}}}_{1}^{1/2})$} &$\Xi^b({\overline{\boldsymbol{3}}}_{0})+\overline{K}$
& 0.813& -- -- --&0.813   \\ 
 & $\Omega^b({\boldsymbol{6}}_{1})+\pi$& 0.079& 0.030& 0.109  \\  
 \cline{2-5} 
&\multicolumn{3}{l|}{~~~total}& {\bf 0.922} \\ \hline
\multirow{2}{*}{$\Omega^b({\boldsymbol{\overline{15}}}_{1}^{3/2})$} &$\Xi^b({\overline{\boldsymbol{3}}}_{0})+\overline{K}$
& 1.027& -- -- --&1.027   \\ 
& $\Omega^b({\boldsymbol{6}}_{1})+\pi$& 0.024& 0.089& 0.113  \\ 
 \cline{2-5} 
&\multicolumn{3}{l|}{~~~total}&{\bf 1.140} \\ \hline
\multirow{2}{*}{$\Xi^b_{3/2}({\overline{\boldsymbol{15}}}_{1}^{1/2})$}& $\Xi^b({\overline{\boldsymbol{3}}}_{0})+\pi$
& 2.599 -- 2.899 & -- -- --&2.599 -- 2.899 \\ 
  & $\Xi^b({\boldsymbol{6}}_{1})+\pi$&0.036 -- 0.042 &0.015 -- 0.018 & 0.051 -- 0.060  \\  
 \cline{2-5} 
&\multicolumn{3}{l|}{~~~total}& {\bf 2.650 -- 2.959} \\ \hline
\multirow{2}{*}{$\Xi^b_{3/2}({\overline{\boldsymbol{15}}}_{1}^{3/2})$}& 
 $\Xi^b({\overline{\boldsymbol{3}}}_{0})+\pi$&2.916 -- 3.237 & -- -- --&2.916 -- 3.237 \\ 
& $\Xi^b({\boldsymbol{6}}_{1})+\pi$
&0.011 -- 0.013 &0.045 -- 0.053 & 0.056 -- 0.0.66  \\
 \cline{2-5} &\multicolumn{3}{l|}{~~~total}& {\bf 2.972 -- 3.303} \\ \hline
 &   $\Xi^b({\boldsymbol{3}}_{0})+\pi$&0.147 -- 0.167 &-- -- --&0.147 -- 0.167\\ 
$\Xi^b_{1/2}({\overline{\boldsymbol{15}}}_{1}^{1/2})$&$\Lambda^b({\overline{\boldsymbol{3}}}_{0})+\overline{K}$
&0.455 -- 0.554 & -- -- --&0.455 -- 0.554 \\ 
 & $\Xi^b({\overline{\boldsymbol{6}}}_{1})+\pi$&0.047 -- 0.058 & 0.020 -- 0.024 &0.067 -- 0.082\\ 
 \cline{2-5} 
&\multicolumn{3}{l|}{~~~total}& {\bf 0.669 -- 0.803}\\ \hline
 \multirow{3}{*}{$\Xi^b_{1/2}({\overline{\boldsymbol{15}}}_{1}^{3/2})$} & $\Xi^b({\boldsymbol{3}}_{0})+\pi$
 &0.166 -- 0.187 &-- -- --& 0.166 -- 0.187\\ 
&  $\Lambda^b({\overline{\boldsymbol{3}}}_{0})+\overline{K}$
&0.549 -- 0.656 & -- -- --&0.549 -- 0.656\\ 
 & $\Xi^c({\overline{\boldsymbol{6}}}_{1})+\pi$&0.014 -- 0.017 & 0.060 -- 0.073& 0.074 -- 0.091  \\ 
 \cline{2-5} 
 &\multicolumn{3}{l|}{~~~total}&{\bf  0.789 -- 0.934} \\ \hline
 \multirow{2}{*}{$\Sigma^b({\overline{\boldsymbol{15}}}_{1}^{1/2})$}&  $\Lambda^b({\overline{\boldsymbol{3}}}_{0})+\pi$
 &1.653 -- 1.796 &-- -- --&1.653 -- 1.796 \\ 
  & $\Sigma^b({\boldsymbol{6}}_{1})+\pi$&0.032 -- 0.037 &0.013 -- 0.015 & 0.045 -- 0.052 \\ 
 \cline{2-5} 
 &\multicolumn{3}{l|}{~~~total}& {\bf 1.698 -- 1.848} \\ \hline
  \multirow{2}{*}{$\Sigma^b({\overline{\boldsymbol{15}}}_{1}^{3/2})$} &  $\Lambda^b({\overline{\boldsymbol{3}}}_{0})+\pi$
  &1.835 -- 1.986&--  -- --&1.835 -- 1.986  \\ 
 & $\Sigma^c({\boldsymbol{6}}_{1})+\pi$&0.010 -- 0.011 &0.040 -- 0.046 &0.049 -- 0.057\\ 
 \cline{2-5} 
 &\multicolumn{3}{l|}{~~~total}& {\bf 1.884 -- 2.043}\\ \hline
 \multirow{2}{*}{$\Lambda^b({\overline{\boldsymbol{15}}}_{1}^{1/2})$}  & $\Lambda^b({\overline{\boldsymbol{3}}}_{0}) + \eta$ 
  & 0.000 -- 0.007  & -- -- -- & 0.000 -- 0.007 \\ 
  & $\Sigma^b({\boldsymbol{6}}_{1})+\pi$
 &0.019 -- 0.028 &0.007 -- 0.012 & 0.026 -- 0.040\\ 
 \cline{2-5} 
 &\multicolumn{3}{l|}{~~~total}&{\bf 0.026 -- 0.047} \\ \hline
 \multirow{2}{*}{$\Lambda^b({\overline{\boldsymbol{15}}}_{1}^{3/2})$}  & $\Lambda^b({\overline{\boldsymbol{3}}}_{0}) + \eta$ 
  & 0.000 -- 0.054  & -- -- -- & 0.000 -- 0.054 \\ 
& $\Sigma^b({\boldsymbol{6}}_{1})+\pi$
&0.006 -- 0.008 &0.023 -- 0.035 & 0.029 -- 0.043 \\ 
 \cline{2-5}
  &\multicolumn{3}{l|}{~~~total}&{\bf 0.029 -- 0.097}\\ \hline
\end{tabular}
\caption{Decay widths of exotic pentaquarks in the SU(3) representation $\overline{\boldsymbol{15}}$ and $J=1$.
Uncertainties correspond to the
mass ranges from Table \ref{tab:mpred} (two 
$\Omega^c$ states are used as the input; therefore, also $\Omega^b$ masses and decay widths are not subject to
such uncertainties).
\label{tab:decs15bar1}}
\end{table*}
\linespread{1}

We compute decay widths of $b\,$-exotica to check whether they follow the pattern
of charm pentaquarks, namely whether the widths of $(\overline{\boldsymbol{15}},J=1)$ are very
small, and the widths $(\overline{\boldsymbol{15}},J=0)$ are large \cite{Praszalowicz:2022hcp}.
We can compute only decays in which the soliton changes its configuration by emitting a pseudo-Goldstone
$\varphi$ boson and the heavy quark is a spectator. Decays into heavy mesons cannot be computed in
the present approach. Nevertheless, we try to estimate approximately their influence on the total widths.

We follow exactly the same steps as in Ref.~\cite{Praszalowicz:2022hcp},
using the decay operator, which
has been obtained via the Goldberger-Treiman relation from the   collective
weak current \cite{Diakonov:1997mm,Kim:2017khv},
\begin{align}
\hat{\mathcal{O}}_{\varphi}&=\frac{1}{2F_{\varphi}}\left[  -\tilde{a}_{1}%
D_{\varphi\,i}^{(8)}-a_{2}\,d_{ibc}D_{\varphi\,b}^{(8)}\hat{J}_{c}-a_{3}%
\frac{1}{\sqrt{3}}D_{\varphi\,8}^{(8)}\hat{J}_{i}\right]  \,p_{i} .  \notag \\
& 
\label{eq:Oai}%
\end{align}
Here $D_{\varphi\,i}^{(8)}$ denote SU(3) Wigner functions, $\hat{J}_i$ is the soliton
spin operator and $p_i$ is the $\varphi$ meson momentum.
Constants $a_{1,2,3}$ that enter Eq.~(\ref{eq:Oai})
were extracted from the semileptonic decays of the baryon octet in
Ref.~\cite{Yang:2015era}:
\begin{equation}
a_{1}\simeq -3.509\,, \;a_{2} \simeq 3.437\, , \; a_{3}   \simeq 0.604\, .
\label{eq:a123}%
\end{equation}
However, due to the fact that $a_1$ scales as $N_{\rm val}$, it has been shown in
Ref.~\cite{Kim:2017khv} that in the heavy quark sector $a_1$ has to be replaced by
\begin{equation}
a_1 \rightarrow \tilde{a}_{1} =-2.1596. \label{eq:a1scaled}%
\end{equation}
With this replacement, all decays of charm and bottom sextet 
and of two exotic $\Omega^c$'s
have been successfully 
described  \cite{Kim:2017khv}.
For the decay constants $F_{\varphi}$, we take
$F_{\pi}=93$~MeV and $F_{K}=F_{\eta}=1.2\,F_{\pi}=112$ MeV.

We compute the decays $B_{1}\rightarrow B_{2}+\varphi$, where $M_{1,2}$
denote masses of the initial and final baryons respectively, and $p_{i}$ is the
c.m. momentum of the outgoing meson of mass $m$~\cite{Kim:2017khv,Diakonov:1997mm}:%
\begin{equation}
\left\vert \vec{p}\, \right\vert =p=\frac{\sqrt{(M_{1}^{2}%
-(M_{2}+m)^{2})(M_{1}^{2}-(M_{2}-m)^{2})}}{2M_{1}} \, .
\end{equation}

The decay width is related to the matrix element of $\hat{\mathcal{O}}_{\varphi}$
squared, summed over the final isospin (but not spin) and averaged over the
initial spin and isospin denoted as $\overline{\left[  \ldots\right]  ^{2}}$;
see the Appendix of Ref.~\cite{Diakonov:1997mm} 
and Erratum of Ref.~\cite{Kim:2017khv}
for the details of the
corresponding calculations,
\begin{equation}
\Gamma_{B_{1}\rightarrow B_{2}+\varphi}=\frac{1}{2\pi}\overline{\left\langle
B_{2}\left\vert \hat{\mathcal{O}}_{\varphi}\right\vert B_{1}\right\rangle ^{2}%
}\,\frac{M_{2}}{M_{1}}p.
\end{equation}

Because $\varphi$ meson is in the SU(3) octet, the following decays are possible:%
\begin{align}
\overline{\boldsymbol{15}}_{J=1}  &  \rightarrow\boldsymbol{6}_{J=1}%
,\overline{\,\boldsymbol{3}}_{J=0},\nonumber\\
\overline{\boldsymbol{15}}_{J=0}  &  \rightarrow\overline{\boldsymbol{15}%
}_{J=1},\boldsymbol{6}_{J=1}.
\label{eq:despatterns}
\end{align}
Direct decays of $\overline{\boldsymbol{15}}_{J=0}$ to the ground state
antitriplet are suppressed. 

After averaging over the initial spin and isospin and summing over the final isospin and over final spin third component
$m_2$, we obtain~\cite{Praszalowicz:2022hcp}:
\begin{align}
\Gamma_{B_{1}\rightarrow B_{2}+\varphi} 
=&  \frac{p^{3}}{24\pi F_{\varphi}^{2}}\frac{M_{2}}{M_{1}} \,
 \frac{\dim\mathcal{R}_{2}}{\dim\overline{\boldsymbol{15}}} 
 \; \gamma_{J_1\rightarrow J_2}(s_{1}\rightarrow s_{2})
  \notag \\
\times &\left( \sum_{\mu}  {\mathcal {G}}_{\overline{\boldsymbol{15}}_{J_1} \rightarrow \mathcal{R}_{2}}^{(\mu)}
\left[
\begin{array}
[c]{cc}%
8 & \mathcal{R}_{2}\\
\varphi & B_{2}%
\end{array}
\right\vert \left.
\begin{array}
[c]{c}%
\overline{\boldsymbol{15}}_{\mu}\\
B_{1}%
\end{array}
\right] \right)^{2} \, .
\label{eq:Gammas}
\end{align}
Here the square bracket stands for the pertinent SU(3) isoscalar factor 
and
$ {\mathcal {G}}^{(\mu)}_{\overline{\boldsymbol{15}}_{J_1}\rightarrow \mathcal{R}_{2}}$ are the decay 
couplings.\footnote{Recall that the soliton in SU(3) $\overline{\boldsymbol{15}}$ can be quantized
as spin $J_1=1$ or $J_1=0$.}
The sum over $\mu$ is relevant only for $\mathcal{R}_{2}=\overline{\boldsymbol{15}}$ in the final 
state. We have~\footnote{Present definitions of the decay constants contain
pertinent SU(3) {\em spin} isoscalar factors, which have not been included in definitions of Ref.~\cite{Kim:2017khv}.} 
\cite{Praszalowicz:2022hcp}
\begin{align}
{\mathcal {G}}_{{\overline{\boldsymbol{15}}_1}\rightarrow {\overline{\boldsymbol{3}}}_0} &=
\sqrt{\frac{1}{2}}
\left(  -\tilde{a}_{1}-\frac
{1}{2}a_{2}\right)
=0.312 \, ,
 \notag \\
{\mathcal {G}}_{\overline{\boldsymbol{15}}_1\rightarrow {\boldsymbol{6}}_1} &=
-\sqrt{\frac{1}{3}}
\left(  -\tilde{a}_{1}-\frac
{1}{2}a_{2}-a_{3}\right)
=0.094\, , 
\label{eq:decayconsts1}
\end{align}
and
\begin{align}
{\mathcal {G}}_{\overline{\boldsymbol{15}}_0\rightarrow {\boldsymbol{6}}_1} &=
\frac{1}{2}
\left(  -\tilde{a}_{1}-\frac
{3}{2}a_{2}\right)
=-1.498\, ,
\notag \\
{\mathcal {G}}^{(\mu=1)}_{{\overline{\boldsymbol{15}}_0}\rightarrow {\overline{\boldsymbol{15}}}_1} &=
\sqrt{\frac{1}{366}}\left(
-\tilde{a}_{1}+\frac{41}{2}a_{2}\right) 
=3.796 \, ,
\notag \\
{\mathcal {G}}^{(\mu=2)}_{{\overline{\boldsymbol{15}}_0}\rightarrow {\overline{\boldsymbol{15}}}_1} &=
-\sqrt{\frac{81}{122}}\left(
-\tilde{a}_{1}+\frac{1}{6}a_{2}\right) 
=- 2.226 \, .
\label{eq:decayconsts0}
\end{align}
Here, we adopt the de Swart conventions for the SU(3) phase factors \cite{deSwart:1963pdg} and label 
the representations as in the numerical code of Ref.~\cite{Kaeding:1995re}.

\linespread{1.28}
\begin{table*}[h]
\centering
\begin{tabular}{|c|c|c|c|c|}
\hline
\multicolumn{2}{|c|}{decay}&\multicolumn{3}{c|}{$\Gamma$~[MeV]}\\
\hline
$B_1$  & $B_2+\varphi$&  $s_2=\frac{1}{2}$ & $s_2=\frac{3}{2}$ & $\Sigma_{s_2}$\\ 
\hline
 & $\Omega^b({\boldsymbol{6}}_{1})+\pi$
 & 38.58 -- 47.41 & 65.23 -- 81.30 &103.81 -- 128.71   \\ 
 $\Omega^b({\overline{\boldsymbol{15}}}_{0})$& $\Xi^b({{\boldsymbol{6}}}_{1})+\overline{K}$
 & 7.41 -- 10.48 & 12.50 -- 18.22& 19.91 -- 28.71  \\ 
 &$\Omega^b({\overline{\boldsymbol{15}}}_{1})+\pi$
 & 0.71 -- 1.88 & 0.69 -- 2.51 & 1.40 -- 4.39   \\ 
 \cline{2-5} 
&\multicolumn{3}{l|}{~~~total}& {\bf 125.12 -- 161.81} \\ \hline
 & $\Xi^b({\boldsymbol{6}}_{1})+\pi$
 &19.46 -- 23.31 &35.49 -- 42.79 & 54.95 -- 66.10  \\ 
 \multirow{2}{*}{$\Xi^b_{3/2}({\overline{\boldsymbol{15}}}_{0})$}& 
 $\Sigma^b({\boldsymbol{6}}_{1})+\overline{K}$
 &16.56 -- 22.11 & 27.41 -- 37.67& 43.97 -- 59.78  \\ 
 &$\Xi^b_{3/2}({\overline{\boldsymbol{15}}}_{1})+\pi$
 &10.40-- 37.18 & 5.19 -- 51.26 & 15.59 -- 88.44 \\ 
 &$\Xi^b_{1/2}({\overline{\boldsymbol{15}}}_{1})+\pi$
 & 1.40 -- 4.31 & 1.54 -- 6.23 &2.95 -- 10.54 \\ 
 \cline{2-5} 
&\multicolumn{3}{l|}{~~~total}& {\bf 117.46 -- 224.86}\\ \hline
 & $\Xi^b({\boldsymbol{6}}_{1})+\pi$
 &25.92 -- 27.42 &46.97 -- 49.80 & 72.88 -- 77.21  \\ 
 \multirow{2}{*}{$\Xi^b_{1/2}({\overline{\boldsymbol{15}}}_{0})$}& 
 $\Sigma^b({\boldsymbol{6}}_{1})+\overline{K}$
 &0.78 -- 0.87 & 1.25 -- 1.40 & 2.03 -- 2.27  \\  
 &$\Xi^b_{3/2}({\overline{\boldsymbol{15}}}_{1})+\pi$
 &0.46 -- 1.78 & 0.00 -- 1.60 & 0.46 -- 3.38   \\ 
 &$\Xi^b_{1/2}({\overline{\boldsymbol{15}}}_{1})+\pi$
 & 0.03  -- 0.09  & 0.02 -- 0.10 & 0.05 -- 0.20 \\ 
 \cline{2-5} 
&\multicolumn{3}{l|}{~~~total}& {\bf 75.43 -- 83.06} \\ \hline
 & $\Sigma^b({\boldsymbol{6}}_{1})+\pi$
 & 14.41 -- 19.13 & 25.46 -- 34.28 & 39.87 -- 53.41 \\ 
 \multirow{2}{*}{ $\Sigma^b({\overline{\boldsymbol{15}}}_{0})$} & $\Sigma^b({\boldsymbol{6}}_{1})+\eta$
 & 0.00 -- 0.12 & -- -- --& 0.00 -- 0.12 \\ 
& $\Sigma^b({\overline{\boldsymbol{15}}}_{1})+\pi$
&0.94 -- 20.26 &0.00 -- 24.56 &0.94 -- 44.82\\ 
 &$\Lambda^b({\overline{\boldsymbol{15}}}_{1})+\pi$
 & 0.10 -- 4.58& 0.00 -- ~6.22& 0.10 -- 10.80\\ 
 \cline{2-5} 
&\multicolumn{3}{l|}{~~~total}& {\bf 40.91 -- 109.15} \\ \hline
 \multirow{2}{*}{$\Lambda^b({\overline{\boldsymbol{15}}}_{0})$} & $\Sigma^b({\boldsymbol{6}}_{1})+\pi$
 &~9.98 -- 11.33 &17.56 -- 20.08 & 27.55 -- 31.41  \\ 
& 
 $\Sigma^b({\overline{\boldsymbol{15}}}_{1})+\pi$
 &0.00 -- 2.20 &  0.00 -- 1.24 &0.00 -- 3.44 \\
 \cline{2-5} 
&\multicolumn{3}{l|}{~~~total}& {\bf 27.55 -- 34.85} \\ \hline
 $N^b({\overline{\boldsymbol{15}}}_{0})$& $N^b({\overline{\boldsymbol{15}}}_{1})+\pi$
 &0.00 -- 18.57 & 0.00 -- 21.03 & 0.00 -- 39.60\\ \cline{2-5} 
 &\multicolumn{3}{l|}{~~~total}& {\bf 0.00 -- 39.60}\\ \hline
\end{tabular}
\caption{Decay widths of exotic $b\;$-pentaquarks in the SU(3) representation $\overline{\boldsymbol{15}}$ and $J=0$.
Uncertainties correspond to the
mass ranges from Table \ref{tab:mpred}.
\label{tab:decs15bar0}}
\end{table*}

Factors $\gamma$ take care of the spin dependence for the soliton angular momenta  $J=1$ or 0
(see Erratum in Ref.~\cite{Kim:2017khv}):%
\begin{align}
\gamma_{1 \rightarrow 1}(1/2 \rightarrow1/2)=2/3,
&\qquad
\gamma_{1\rightarrow 1}(1/2\rightarrow3/2)=1/3, 
\notag \\
\gamma_{1\rightarrow 1}(3/2 \rightarrow1/2)=1/6,
&\qquad
\gamma_{1\rightarrow 1}(3/2\rightarrow3/2)=5/6, 
\notag \\
\gamma_{0 \rightarrow 1}(1/2\rightarrow1/2)=1/3, 
&\qquad
\gamma_{0 \rightarrow 1}(1/2\rightarrow3/2)=2/3, 
\notag \\
\gamma_{1 \rightarrow 0}(1/2\rightarrow 1/2)=1,~~\,
&\qquad
\gamma_{1 \rightarrow 0}(3/2\rightarrow 1/2)=1.
\end{align}
Note that
\begin{equation}
\sum_{s_2} \gamma_{J_1 \rightarrow J_2}(s_1\rightarrow s_2)=1\, .
\end{equation}

\begin{table}[h!]
\centering
\begin{tabular}[c]{|c|c|c|c|}
\hline
$\overline{\boldsymbol{15}}_J$ & final & $J=1$ & $J=0 $\\\hline
$\Omega^{b}$ & $\Xi+B$ & no & yes\\\hline
$\Xi^{b}_{3/2}$ & $\Sigma+B$ & no & yes\\\hline
$\Xi^{b}_{1/2}$ & $\Xi+B_{s}$ & no & no\\
& $\Lambda+B $ & no & yes\\
& $\Sigma+B$ & no & no\\\hline
$\Sigma^{b}$ & $\Sigma+B_{s} $ & no & no\\
& $N+B$ & no & yes\\\hline
$\Lambda^{b}$ & $\Lambda+B_{s}$ & no & no\\
& $N+B$ & no & yes\\\hline
$N^{b}$ & $N+B_{s}$ & no & no\\\hline
\end{tabular}
\caption{Decays of $\overline{\boldsymbol{15}}_{J=1,0} $ to the ground state
octet baryons and $B$ mesons (there is no distinction between particles and antiparticles
since only the masses matter for thresholds).  }%
\label{tab:decs15bartoB}%
\end{table}

We see from Eqs.~(\ref{eq:decayconsts1}) that decay constants of $(\overline{\boldsymbol{15}},J=1)$ are very small;
in fact they vanish in the large $N_c$ limit \cite{Praszalowicz:2018upb}. On the contrary, decay constants of 
$(\overline{\boldsymbol{15}},J=0)$ are almost an order of magnitude larger (\ref{eq:decayconsts0}), so we expect the corresponding decay widths
to be large (also the phase space factor  is larger than in the $J=1$ case).

Numerical results  are presented in Tables~\ref{tab:decs15bar1} and \ref{tab:decs15bar0}.
Decays not included in the tables are not allowed by energy or isospin conservation. Note that these widths are pure 
predictions based on the light sector va{\-}lu{\-}es of the decay parameters, except for rescaling (\ref{eq:a1scaled}). 
In Tab.~\ref{tab:decs15bartoB} we list all possible strong decays to heavy mesons and indicate whether they are
allowed by energy conservation.

Table~\ref{tab:decs15bar1} shows that all states in  $\overline{\boldsymbol{15}}_{J=1}$, have very small
widths, in most cases not exceeding 2~MeV. For all states in $\overline{\boldsymbol{15}}_{J=1}$, decay channels 
to light baryons and heavy mesons are closed.
We therefore conclude that exotic $b\,$-pentaquarks from $\overline{\boldsymbol{15}}_{J=1}$ can be found only in dedicated searches 
in high resolution experiments.
Interestingly, the lightest
member of $\overline{\boldsymbol{15}}_{J=1}$, namely the nucleonlike pentaquark, is stable with respect to two body strong
decays.

The situation is very different in the case of $\overline{\boldsymbol{15}}_{J=0}$ widths listed in Table~\ref{tab:decs15bar0}.
Here all decay widths 
are within 30~--~220~MeV range. The only exception is again the lightest nucleonlike pentaquark, which 
can decay only to the $N^b$ state in $\overline{\boldsymbol{15}}_{J=1}$, which is stable against strong decays.
Furthermore, as seen from Tab.~\ref{tab:decs15bartoB},
all states in $\overline{\boldsymbol{15}}_{J=0}$ (except for $N^b$) have one open channel to the decays
to light baryons and heavy mesons. We are not able to compute these widths within the present approach.
However, since the available phase space is comparable to the decays listed in Tab.~\ref{tab:decs15bar0},
we may expect that the total decay widths will double with respect to the estimates given in Tab.~\ref{tab:decs15bar0}.

One should also note, that all decays of $\overline{\boldsymbol{15}}_{J=0}$ lead to either $\boldsymbol{6}$
or $\overline{\boldsymbol{15}}_{J=1}$, which decay further to $\boldsymbol{6}$ and $\overline{\boldsymbol{3}}$.

We conclude therefore, that pentaquarks from $\overline{\boldsymbol{15}}_{J=0}$ multiplet are very wide and 
-- similarly to the charm case discussed in Ref.~\cite{Praszalowicz:2022hcp} -- may be
interpreted as a background, rather than as a signal. Therefore, they could have been missed in general purpose experiments.

\section{Summary and conclusions}
\label{sec:conclusions}

In the present paper we have closely followed reference \cite{Praszalowicz:2022hcp} on charm $\overline{{\bf 15}}$ pentaquarks extending 
it to the case of beauty. We have used the $\chi$QSM where all model parameters were extracted from the
the heavy baryon spectra including charm exotica and from the negativity of splitting
parameters $\alpha,~\beta$, and $\gamma$ (\ref{eq:Hsb}). In this way we in fact tested the underlying {\em hedgehog} SU(3) symmetry 
 rather  than
a particular implementation of the model. We did not
present numerical support for the model mass formulas (\ref{eq:15barav}), (\ref{eq:Delta15bar}) and (\ref{eq:M3barM6mass})
and other regularities that were at length discussed in Ref.~\cite{Praszalowicz:2022hcp}.

We computed masses of $b$ pentaquarks
with uncertainties of the order $\sim50$~MeV, which are grouped in three $\overline{{\bf 15}}$ multiplets
spanned in a range 6000 -- 6600~MeV. Question arises, can one observe these states experimentally?

It was claimed in  Ref.~\cite{Praszalowicz:2022hcp} the two LHCb heavy $\Omega^{c 0}$ states \cite{LHCb:2017uwr,Belle:2017ext,LHCb:2021ptx}
could be
interpreted as charm pentaquarks due to their extremely small widths. Unfortunately the $\Omega^{b -}$
states reported in Ref.~\cite{LHCb:2020tqd} are much lighter than the predictions of the present approach
(even lighter than possible regular ${\bf 6}$ states). Moreover, their mass differences that are much smaller than the 
$\Omega^c$ mass splittings in the charm sector, which is strange since the mass splittings in all know cases
are basically independent of the heavy quark mass.

Finally, we computed  strong decay widths using the same approach as Ref.~\cite{Praszalowicz:2022hcp}.
We have found that the decay pattern is very similar to the charm case.
Pentaquarks belonging to the $\overline{\boldsymbol{15}}_{J=1}$ SU(3) multiplet
are very narrow having widths of the order of $\sim 2$~MeV, while the remaining states from the
$\overline{\boldsymbol{15}}_{J=0}$ SU(3) multiplet are wide, in most cases of the order of $\sim 100$~MeV or more.
Moreover, all these decays lead to the unstable resonances.
 For all pentaquarks in $\overline{\boldsymbol{15}}_{J=1}$ decay channels to heavy mesons are closed, and on
the contrary, $\overline{\boldsymbol{15}}_{J=0}$ have open channels to the final states containing $B$ mesons,
which can double decay widths computed in the $\chi$QSM.
Multipurpose searches can easily miss narrow or wide exotic states, so dedicated studies are needed to confirm/reject
our predictions.

\section*{Acknowledgments}
This work has been supported by the Polish National Science Centre Grants No. 2017/27/B/ST2/01314.
The author acknowledges CERN TH Department for hospitality while this research was being carried out.

\end{document}